\documentclass[a4paper,11pt]{article}
\usepackage{pos}

\title{Mitigating the Hubbard Sign Problem. A Novel Application of Machine Learning}
\author*[a,b,c,d]{Marcel Rodekamp} 
\author[a,c,d]{Christoph Gäntgen}

\newcommand{\bonn}{
    Helmholtz-Institut f\"{u}r Strahlen- und Kernphysik,
    Rheinische Friedrich-Wilhelms-Universit\"{a}t Bonn, 53115 Bonn, Germany
}

\newcommand{\ias}{
    Institute for Advanced Simulation,
    Forschungszentrum J\"{u}lich, 54245 J\"{u}lich, Germany
}

\newcommand{\jsc}{
    JARA \& J\"{u}lich Supercomputing Center,
    Forschungszentrum J\"{u}lich, 54245 J\"{u}lich, Germany
}

\newcommand{\casa}{
    Center for Advanced Simulation and Analytics (CASA),
    Forschungszentrum Jülich, 52425 J\"{u}lich, Germany
}
 \affiliation[a]{\ias}
\affiliation[b]{\jsc}
\affiliation[c]{\casa}
\affiliation[d]{\bonn}

\emailAdd{m.rodekamp@fz-juelich.de}
\emailAdd{c.gaentgen@fz-juelich.de}

\abstract{
Many fascinating systems suffer from a severe (complex action) sign problem preventing us from calculating them with Markov Chain Monte Carlo simulations. One promising method to alleviate the sign problem is the transformation of the integration domain towards Lefschetz Thimbles. Unfortunately, this suffers from poor scaling originating in numerically integrating of flow equations and evaluation of an induced Jacobian. In this proceedings we present a new preliminary Neural Network architecture based on complex-valued affine coupling layers. This network performs such a transformation efficiently, ultimately allowing simulation of systems with a severe sign problem. We test this method within the Hubbard Model at finite chemical potential, modelling strongly correlated electrons on a spatial lattice of ions.
}
 
\FullConference{%
  The 39th International Symposium on Lattice Field Theory (Lattice2022),\\
  8-13 August, 2022 \\
  Bonn, Germany 
}

\usepackage[T1]{fontenc}
\usepackage[utf8]{inputenc}
\usepackage[UKenglish]{babel}

\usepackage{pgfplots}
\usepackage{tikz}
\usetikzlibrary{
    arrows,
    arrows.meta,
    shapes,
    decorations,
    decorations.shapes,
    calc
}
\usepackage{ifthen}

\usepackage{physics}
\usepackage{siunitx}
\usepackage[super]{nth}
\usepackage{dsfont}
\usepackage{nicefrac}
\usepackage{mathtools}
\usepackage{amsmath}

\usepackage{booktabs}

\usepackage{hyperref}

\usepackage{xcolor}
\usepackage{tcolorbox}
\tcbuselibrary{theorems}
\tikzstyle{arrow} = [very thick,->,>=stealth]

\tikzset{
parallel segment/.style={
   segment distance/.store in=\segDistance,
   segment pos/.store in=\segPos,
   segment length/.store in=\segLength,
   to path={
   ($(\tikztostart)!\segPos!(\tikztotarget)!\segLength/2!(\tikztostart)!\segDistance!90:(\tikztotarget)$) -- 
   ($(\tikztostart)!\segPos!(\tikztotarget)!\segLength/2!(\tikztotarget)!\segDistance!-90:(\tikztostart)$)  \tikztonodes
   }, 
   segment pos=.5,
   segment length=4ex,
   segment distance=1mm,
},
}

\definecolor{fzjblue}{RGB}{2,61,107} %
\definecolor{fzjlightblue}{RGB}{173,189,227} %
\definecolor{fzjgray}{RGB}{235,235,235} %
\definecolor{fzjred}{RGB}{235, 95, 115}  %
\definecolor{fzjgreen}{RGB}{185, 210, 95}  %
\definecolor{fzjyellow}{RGB}{250, 235, 90}  %
\definecolor{fzjviolet}{RGB}{175, 130, 185}  %
\definecolor{fzjorange}{RGB}{250, 180, 90}  %
\definecolor{fzjblack}{RGB}{0,0,0}
\definecolor{fzjwhite}{RGB}{255,255,255}

\newcommand{\Ncfg}{\ensuremath{\mathrm{N}_{\mathrm{conf}}}}
\newcommand{\Nt}{\ensuremath{\mathrm{N}_{\mathrm{t}}}}
\newcommand{\Nx}{\ensuremath{\mathrm{N}_{\mathrm{x}}}}
\newcommand{\Obs}{\ensuremath{\mathcal{O}}}
\newcommand{\Z}{\ensuremath{\mathcal{Z}}}
\newcommand{\M}{\ensuremath{\mathcal{M}}}
\newcommand{\Reals}{\ensuremath{\mathbb{R}}}
\newcommand{\Complexs}{\ensuremath{\mathbb{C}}}
\newcommand{\DD}[1]{\ensuremath{\hspace*{-3pt}\mathcal{D}#1\;}}

\newcommand{\SHIFT}{\ensuremath{\mathrm{SHIFT}}}
\newcommand{\Seff}{\ensuremath{S_{\mathrm{eff}}}}
\newcommand{\NN}{\ensuremath{\mathcal{NN}}}
 
\begin{document}
\maketitle

\section{Introduction}

Markov Chain Monte Carlo algorithms (MCMC) enjoy great success in simulating 
many theories from the Ising model up to Lattice QCD. 
Albeit the potential, MCMC has a hard time whenever the action 
becomes complex-valued due to the associated Boltzmann weight loosing its 
interpretability as probability distribution. 

Using MCMC, we focus on the Hubbard model capturing electronic properties 
of systems with strongly interacting electrons propagating on a fixed 
spatial lattice of ions.
Examples for such systems are carbon nano structures like Graphene and 
Fullerene $C_n$.
In the Hubbard model the sign problem is observed at finite chemical potential as well as on 
non-bipartite lattices\footnote{ 
    Bipartite describes lattice geometries at which the sites can be two 
    coloured such that no neighbouring sites have the same colour.
    For example, the square is bipartite while the triangle is non-bipartite.
}.
Reweighting can treat the complex-valued Boltzmann weight though, at the 
same time, introducing an exponentially hard to estimate normalization.

Deforming the region of integration onto Manifolds with an almost constant imaginary 
action showed great promise in reducing the sign problem substantially~\cite{
Kashiwa:2018vxr, alexandru2020complex, DetmoldPathIntegral, Detmold:2021ulb}.
Practically, this deformation requires numerical integration of differential 
equations which becomes infeasible for larger systems. 
We aim to identify efficient Neural Network architectures to learn such  
beneficial deformations.
This removes the cost of numerically integrating configurations and enables 
simulations of large systems with a sign problem beyond the standard reweighting
approach.

In this proceedings, we collect material from our earlier publications~\cite{leveragingML,rodekampMitigating} and a master thesis~\cite{christophThesis}.
The manuscript is organized in the following way.
In section~\ref{sec:Formalism} a brief introduction to the Hubbard model and 
the tested system is presented along a short discussion of the sign problem.
This discussion is then followed by the definition of the neural network 
architectures as published in~\cite{leveragingML,rodekampMitigating}.
Further, in section~\ref{sec:NumericalResults} correlator results are presented 
and we discuss the obtained charge density of one of the larger systems.
\label{sec:Introduction}

\section{Formalism}\label{sec:Formalism}
\begin{figure}
    \centering
    \includegraphics[scale=1] {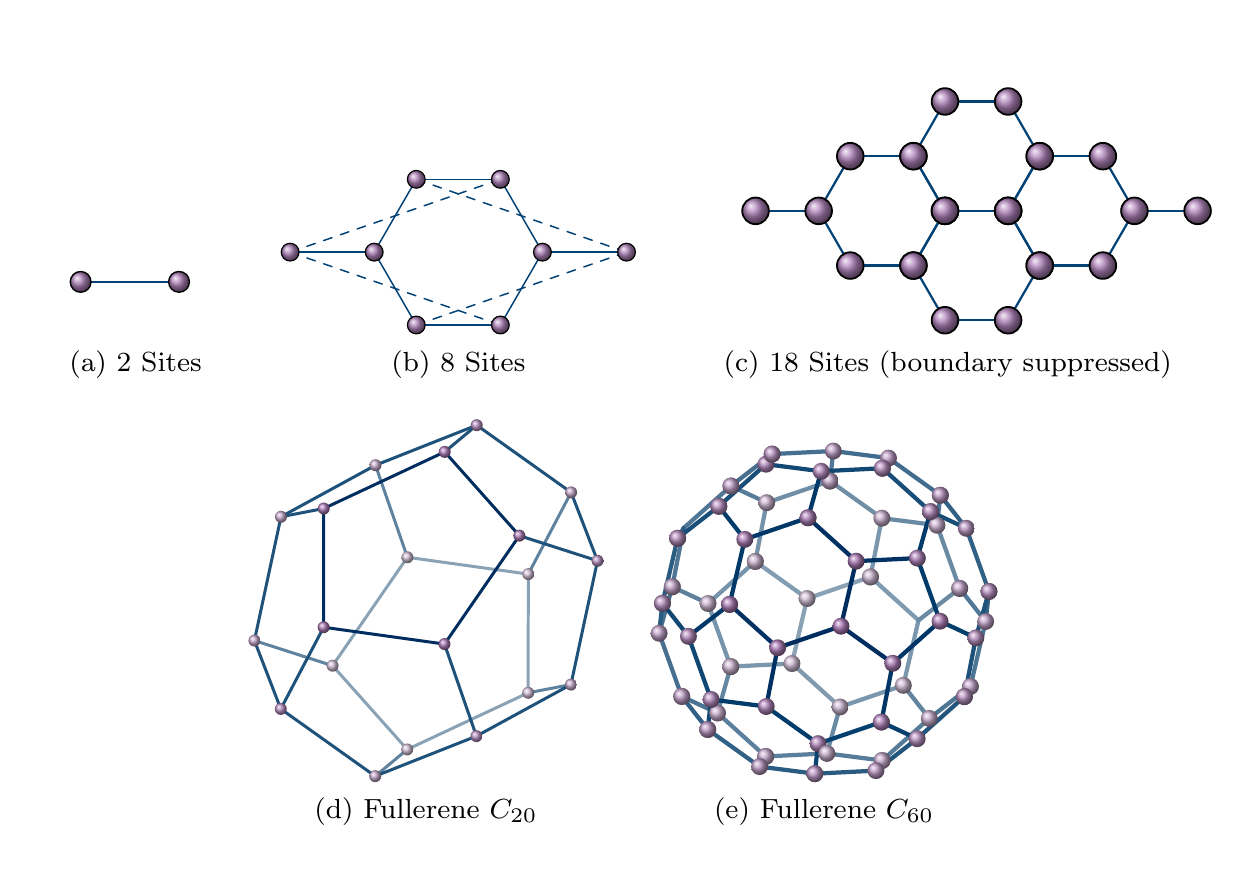}
    \caption{
        Showing the different geometries considered in this proceedings. 
        Each vertex represents an ion and each (dashed) line depicts the 
        nearest-neighbor hopping that is allowed by the Hubbard model. 
        Dashed lines indicate periodic boundary condition where possible.
    }
    \label{fig:systems}
\end{figure}

The Hubbard model~\cite{Hubbard1963} describes a fixed spatial lattice $X$ of ions on which electrons can move and interact.
Its Hamiltonian, in particle-hole basis, is
\begin{equation}
    \nonumber
    \mathcal{H}\left[K, V, \mu\right]
    =
           - \sum_{x,y\in X} \left( p_x^\dagger K^{xy} p_y - h_x^\dagger K^{xy} h_y \right)
         + \frac{1}{2} \sum_{x,y\in X} \rho_x V^{xy} \rho_y
            + \mu \sum_{x\in X} \rho_x,
    \label{eq:hubbard-hamiltonian}
\end{equation}
where the amplitudes in $K$ encode the hopping of fermionic particles $p$ 
and holes $h$, the potential $V$ describes the interactions between charges
\begin{equation}
    \rho_x = p^{\dagger}_x p_x - h^{\dagger}_x h_x
    \label{eq:net-charge}
\end{equation}
and the chemical potential $\mu$ incentivizes charge.
By adjusting the hopping and lattice symmetry $K$ as well as the interaction $V$ 
this model can describe a wide variety of physical systems.
In the following investigation, five systems are considered as displayed in figure~\ref{fig:systems}.
The 2 site system describes one unit cell of the honeycomb structure used 
for Graphene type models which we successively built up with the 8 and 18 site ones.
Secondly, we present preliminary results for fullerenes $C_{20}$ and $C_{60}$ at zero chemical potential.
In all cases $K$ encodes nearest-neighbor hopping
and we assume an on-site interaction,
\begin{align}
    K &= \kappa \delta_{\langle xy \rangle}
    &
    V &= U \delta_{xy}.
\end{align}
In figure~\ref{fig:systems} the sites, i.e.\@ ions and their nearest neighbor connections are depicted. 
Lines stretching out display periodic boundary of the spatial lattice (suppress in the 18 site case).

\subsection{Simulation Setup}
Calculating observables follows the standard procedure~\cite{browerHybridMonteCarlo2012} 
of evaluating the thermal trace.
After trotterizing into $\Nt$ time slices, inserting Grassmannian resolutions of
the identity and linearizing the interaction via the Hubbard-Stratonovich 
transformation~\cite{Hubbard1959} the Hamiltonian is transformed into the action
\begin{equation}
    S\left[\Phi \,\vert \, K, V, \mu \right]
       =
            -   \log\det{ M\left[\Phi\,\vert\, K,\mu\right] \cdot M\left[-\Phi\,\vert\, -K,-\mu\right] }
         +   \frac{1}{2} \sum_{t=0}^{\Nt-1}\sum_{x,y\in X} \Phi_{tx} \delta V^{-1\,xy} \Phi_{ty},
    \label{eq:hubbard-action}
\end{equation}
where $\Phi \in \Reals^{\abs{\Lambda}}$ is an auxiliary field on the 
spacetime lattice $\Lambda = [0, N_t-1]\otimes X$ and $\delta=\beta/N_t$ 
is the (temporal) lattice spacing controlled by the inverse temperature $\beta$.
The fermion matrix is not uniquely defined on the lattice, we choose the exponential discretization~\cite{Wynen:2018ryx}
\begin{equation}
    M\left[\Phi\,\vert\, K,\mu\right]_{x't';xt} 
    =   
        \delta_{x'x}\delta_{t't}
    -  \left( e^{\delta(K + \mu)} \right)_{x'x} e^{+ i \Phi_{xt}} \mathcal{B}_{t'}\delta_{t'(t+1)}
\end{equation}
where $\mathcal{B}$ encodes the anti-periodic boundary conditions in time.
For bipartite systems we may replace $-K$ with $+K$ in the holes' fermion 
matrix~\cite{browerHybridMonteCarlo2012}.
The thermal trace for this is expressed as path integral
\begin{equation}
    \expval{\Obs}
    =
    \frac{1}{\Z} \int  \DD{\Phi} e^{- S\left[\Phi\right]} \Obs\left[\Phi\right]
    =
    \int  \DD{\Phi} p_S\left[\Phi\right] \Obs\left[\Phi\right]
    \label{eq:true-expectation-value}
\end{equation}
For cases of real action we can apply MCMC to generate $\Ncfg$ configurations according to the Boltzmann distribution $p_S\left[\Phi\right] $
and estimate observables \eqref{eq:true-expectation-value} by an unweighted expectation value.
If the action is complex valued we apply reweighting
\begin{align}
        \expval{\Obs}
        &   = \frac{\expval{e^{-i\Im{S}}\Obs}_{\Re{S}}}{\expval{e^{-i \Im{S}}}_{\Re{S}}}
            = \frac{1}{\Sigma} \expval{ e^{-i \Im{S}} \Obs}_{\Re{S}}.
        \label{eq:reweighting}
\end{align}
and sample configurations according to the Boltzmann distribution under the real part of the action.
It has been shown~\cite{berger2021complex} that an effective number of configurations
\begin{equation}
    \Ncfg^{\textrm{eff}} = \abs{\Sigma}^2 \cdot \Ncfg \label{eq:effective-Nconf}
\end{equation}
controls the scaling of statistical errors $\sim \left(\Ncfg^{\textrm{eff}}\right)^{-1/2}$.
This translates the sign problem to the ability of calculating the denominator $\Sigma$, 
i.e. the statistical power, reliably~\cite{berger2021complex,leveragingML,PhysRevD.93.014504,mori2018lefschetz}.
\\

A promising approach to mitigate, or even remove, the sign problem is to 
deform the region of integration $\Phi \in \M_{\Reals} = \Reals^{\abs{\Lambda}}$ onto a manifold 
on which the imaginary part of the action is (nearly\footnote{
    If $\Im{S}\approx const$, the statistical power $\abs{\Sigma} = 
    \abs{\expval{e^{i\Im{\Seff}}}} \approx \abs{\expval{e^{i \cdot const}}} \approx 1$
    yielding nearly no reduction in effective number of configurations 
    $\Ncfg^{\mathrm{eff}} \approx \Ncfg$.
}) constant~\cite{PhysRevD.86.074506,alexandru2020complex}
, $\M = \left\{ \Phi \in \Complexs^\abs{\Lambda} \, \vert \, \Im {S\left[\Phi\right]} = const. \right\}$.
If $\M$ is in the same homology class as $\M_{\Reals}$
an analogue of the Cauchy integral theorem ensures that the observables are unchanged.
Parametrizing fields $\Phi(\phi)\in\M$ then adds a Jacobian defining the 
used effective action
\begin{equation}
    \Seff\left[\phi\right] = S[\Phi\left(\phi\right)] - \log\det{J\left[\phi\right]}, \quad J_{ij} = \pdv{\Phi(\phi)_i}{\phi_j}
    \label{eq:effective-action}
\end{equation}

There are many strategies for picking target manifolds $\M$~\cite{Tanizaki:2017yow}.
One choice is to try to approximate the Lefschetz thimbles~\cite{PhysRevD.86.074506}.
Each thimble contains a critical point $\Phi_{cr}$ that is a fixed point 
of the holomorphic flow
\begin{equation}
    \frac{d\Phi(\tau)}{d\tau} = \pm\left(\frac{\partial S\left[ \Phi(\tau) \right]}{\partial \Phi(\tau)}\right)^*, \quad \Phi(0) = \phi
    \label{eq:holomorphic-flow}
\end{equation}
introducing the fictitious flow time $\tau$.
A thimble is the set of complexified configurations that flow to a critical 
point under downward flow, i.e. $-$ in~\eqref{eq:holomorphic-flow}.
An integrator for~\eqref{eq:holomorphic-flow} will always be computationally 
expensive~\cite{alexandru2020complex,rodekampMitigating,leveragingML}.
However, the non-interacting solution $\phi = 0$ for~\eqref{eq:holomorphic-flow}
assuming a constant field $\Phi_{t,x} = \Phi_{t',x'}$ defines 
a (hyper-) plane parallel to the real plane $\M_\Reals$ that goes through the main critical point $i \Phi_c^0$.
This so called tangent plane $\M_{T} = \{\phi + i \Phi_c^0 \, \vert \, \forall \phi \in \M_{\Reals}\}$
showed promise in sufficiently mitigating the sign problem, at least for smaller systems~\cite{PhysRevD.93.014504,Alexandru:2018ddf, Warrington:2019kzf, leveragingML, rodekampMitigating}.

\subsubsection{Neural Network Architectures}
To improve beyond the tangent plane it seems plausible to 
identify a transformation that transforms a given configuration $\phi\in\M_\Reals$ 
to a target manifold $\tilde{\M}$ that may be closer to $\M$ than the tangent plane.
Such a transformation may be parametrized by a neural network 
\begin{equation}
    \SHIFT: \M_\Reals \to \tilde{\M}, \, \phi \mapsto \phi + i\left(\Phi_c^0 + NN\left( \phi \right)\right).
    \label{eq:SHIFT}
\end{equation}
For the neural network part $NN$ we pick two linear layers $\omega \phi + b$ with real trainable weights $\omega$ and biases $b$
which are separated by a leacky-ReLU. 
As the effective action~\eqref{eq:effective-action} suggests the defining 
transformations Jacobian needs to be computed 
\begin{equation}
    \log\det{J_{\SHIFT}[\phi]} = \log\det{\mathds{1} + i \pdv{NN\left(\phi\right)}{\phi} } 
\end{equation}
which requires an $\order{V^3}$ algorithm for the determinant calculation.
This scaling is not feasible for large scale systems but still much cheaper 
than applying a Runge-Kutta --- or similar algorithms --- to integrate the 
holomorphic flow equations.

To improve the scaling, we identify a neural network that has a cheaper determinant. 
One such architectures is given by Affine Coupling Layers~\cite{albergo2021introduction,foreman2021hmc}
that approximate the integrator $\Phi(\phi) \approx \NN(\phi) $
\begin{equation}
    \NN_\ell(\Phi) =
    \begin{cases}
        c_\ell\left[ \Phi_A, \, \Phi_B \right] & A_\ell \text{ components} \\
        \Phi_B & B_\ell \text{ components}
    \end{cases}
    \label{eq:ACL-def}
\end{equation}
Here $A$ and $B$ are layer-specific partitions of the input vector $\Phi$ of equal cardinality $\nicefrac{1}{2}\abs{\Lambda}$. $\Phi_{A,B}$ are the components of the input belonging to the indicated partition.
We apply the affine coupling~\cite{albergo2021introduction}
\begin{equation}
    c_\ell\left[\Phi_A, \Phi_B \right] = e^{m_\ell\left(\Phi_B\right)} \odot \Phi_A + a_\ell\left(\Phi_B\right).
    \label{eq:affine-coupling}
\end{equation}
The coupling functions 
$m_\ell,a_\ell: \mathbb{C}^{\nicefrac{\abs{\Lambda}}{2}} \to \mathbb{C}^{\nicefrac{\abs{\Lambda}}{2}}$ 
are again two linear layers separated by the non-linear Softsign function.
To ensure that the neural network produces a complex configuration as is required by the holomorphic flow, 
the weights and biases need to be complex valued which is discussed in more detail in~\cite{rodekampMitigating}.
A single layer just transforms half of the configuration, we thus pair it up with a second layer that is set up in the same way but with the roles of $A$ and $B$ interchanged.
This setup allows to express the Jacobian, with $L/2$ of these pairs, in the form
\begin{equation}
    \log{\det{J_{\NN}(\phi)}} = \sum_{\ell = 1}^{L} \sum_{i =0}^{\abs{A}-1} m_\ell\left( \Phi_{\ell-1}(\phi)_B\right)_i.
    \label{eq:logDetJ-NN}
\end{equation}
which adds only an $\order{V}$ cost to the application of the transformation.

For any of these architectures we perform Molecular Dynamics on $\M_\Reals$ using 
a standard leapfrog algorithm and then apply the network to move onto $\tilde{\M}$.
Accept/Reject is then performed according to the effective action~\eqref{eq:effective-action}
using the transformed configuration from the Network and the Jacobian defined by the 
network.
This machine learning enhanced Hybrid Monte Carlo is referred to as ML HMC.

\subsection{Observables}
We are interested in the electronic properties of a given system.
Euclidean correlation functions of a single particle or a single hole %
\begin{equation}
    C_{xy}^{\tiny\substack{p\\h}}(t) = \expval{ p_x^\dagger(t) p_y(0) } = \expval{ M[\pm\Phi | \pm K, \pm\mu ]^{-1}_{xt;y0} },
\end{equation}
momentum projected and averaged we obtain $C_{sp}(t)$~\cite{rodekampMitigating}.
In the future we aim to match the parameters $\nicefrac{U}{\kappa},\mu$ to real-world systems
and extract the band-gap $C_{xy}(t) \sim e^{-t\cdot \Delta E} $.
Further, the charge density is defined by
\begin{equation}
    \rho(\mu) = \frac{1}{\abs{X}} \sum_{x\in X} \expval{\rho_x} 
            = \frac{1}{\abs{X}} \sum_{x\in X} C_{x,x}^p(0) - C_{x,x}^h(0).
            \label{eq:charge-density}
\end{equation}
It is point symmetric around the electric neutral half-filling point, $\mu = 0$, due to 
particle-hole symmetry.
For very large $\mu\to \pm \infty$ the charges (+) or holes (-) are favoured
yielding a charge density of $\pm 1$.
Qualitatively, it is expected that the system's charge 
equals integer multiples of the electric charge $n\cdot e^{-}$ with 
$n \in [-\Nx,\Nx]$, i.e.\@ $\rho(\mu) = \nicefrac{n}{\Nx}$.

\section{Numerical Results}\label{sec:NumericalResults}
We experimented with different training setups. 
Foremost, supervised training using ADAM to minimize the $L1-$Loss.
The training data consists out of $\order{\num{10000}}$ pairs $(\phi,\Phi(\phi))$
obtained by a Runge-Kutta of \nth{4} order.
For further details consider~\cite{leveragingML,rodekampMitigating}.

\begin{figure}
\centering
\includegraphics[width=0.9\linewidth,height = 0.4\textheight]{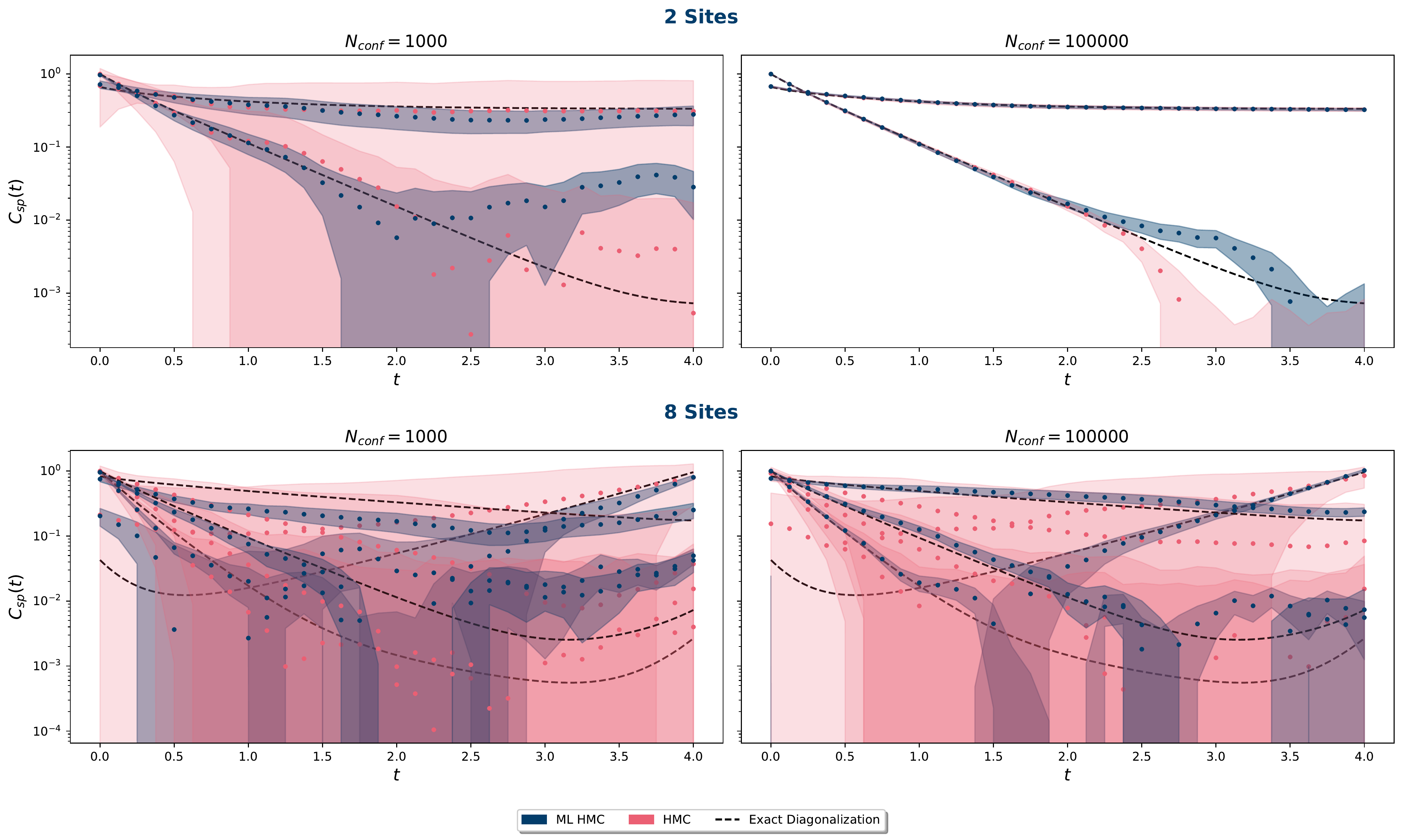}
\caption{
    Each row in this figure shows the correlators
    of a system with 2 (upper row) and 8 (lower row) sites~\cite{rodekampMitigating}.
    The different columns correspond to a number of configurations, 
    $\Ncfg\in\{\num{1000},\num{100000}\}$, used to estimate the correlators.
    Three methods --- ML HMC with coupling layer $\NN$ (blue), HMC on the Tangent Plane (red) and exact diagonalization (black) --- 
    are compared to show the effectiveness and correctness of the introduced 
    machine learning enhanced method. 
    The sign problem manifests as a loss of signal, 
    i.e. small number of effective configurations $\Ncfg^{\mathrm{eff}}$~\eqref{eq:effective-Nconf}, 
    and greatly increases as the number of sites expands.
    It can be seen that the ML HMC has a substantially reduced sign problem.
}
\label{fig:SmallSystemCorrelators}
\end{figure}
\begin{figure}
\centering
\includegraphics[width=0.9\linewidth,height = 0.3\textheight]{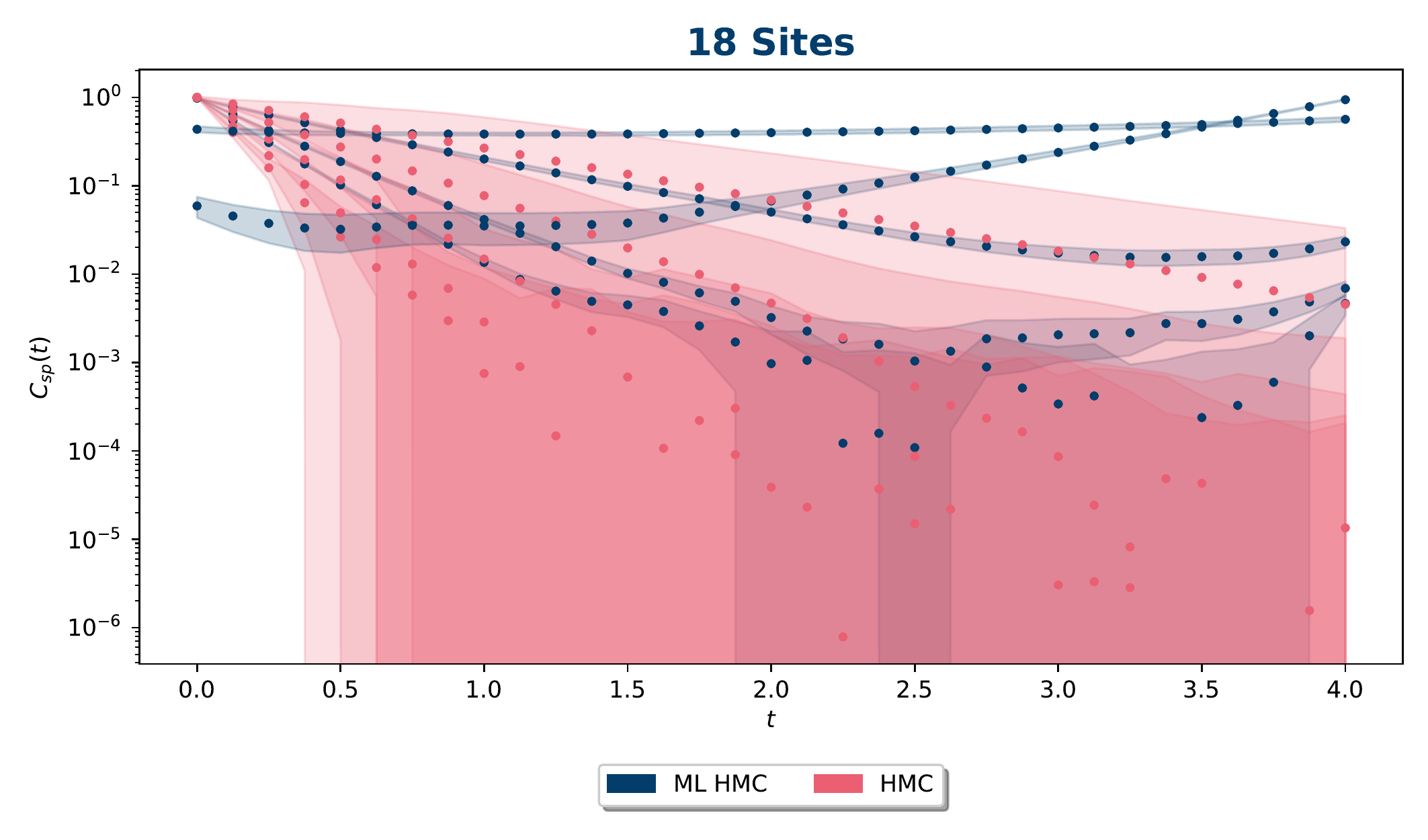}
\caption{The correlators of Graphene with 18 ions are shown~\cite{rodekampMitigating}.
    $\num{100000}$ measurements are taken.
    With this larger lattice direct diagonalization as in figure~\ref{fig:SmallSystemCorrelators} 
    is not tractable any more hence only the two statistical methods ML HMC using the coupling network $\NN$ (blue)
    and HMC on the tangent plane (red) are compared. 
    As expected the ML HMC improves the signal drastically.
}
\label{fig:18SiteCorrelator}
\end{figure}
\begin{figure}
    \centering
\includegraphics[width = 0.9\linewidth,height = 0.4\textheight]{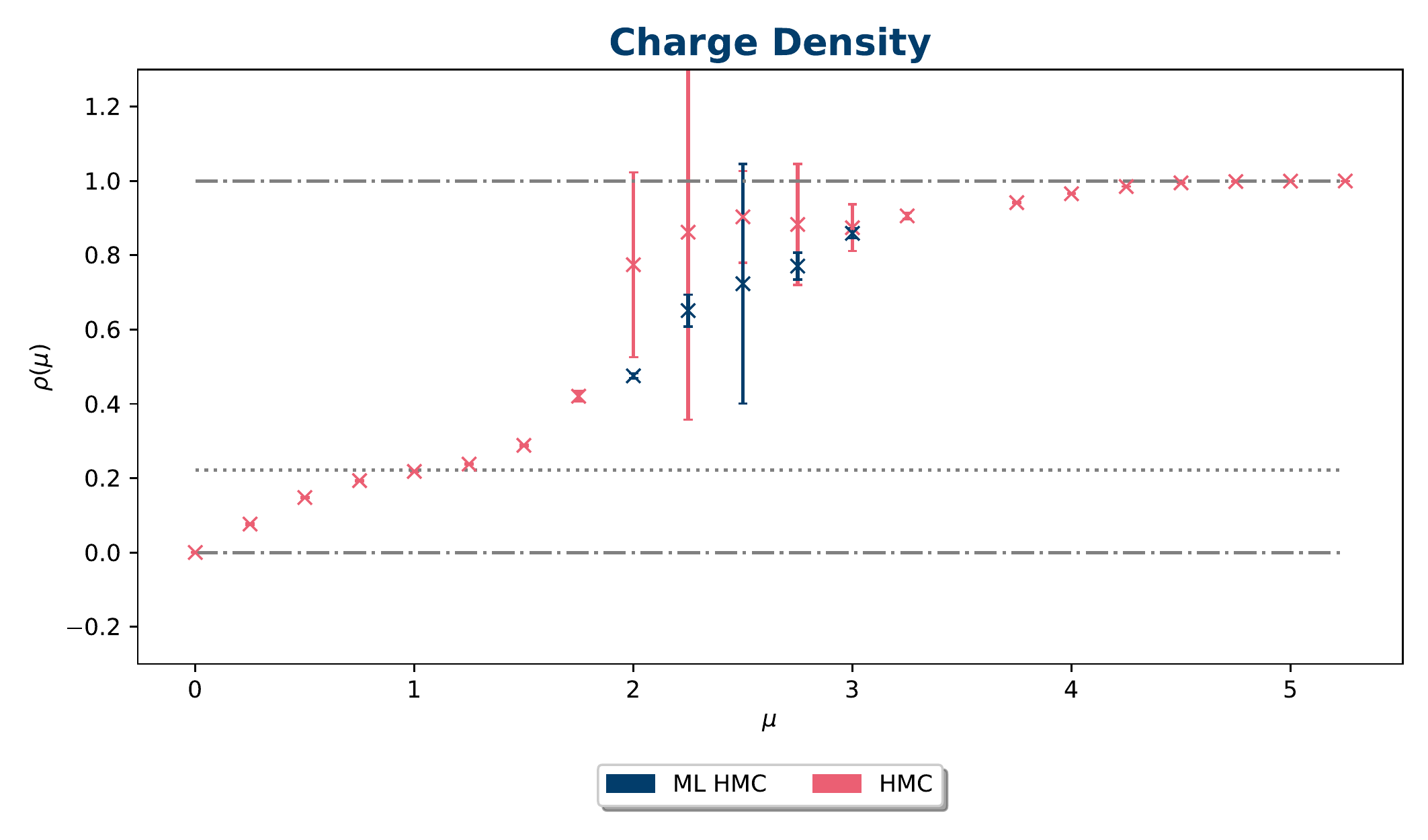}
\caption{
    We computed the charge density for several values of the chemical potential $\mu\in \left[0,\,5.2\right]$
    for an 18 site Graphene sheet~\cite{christophThesis}.
    For most smaller and larger values of $\mu$ the sign problem is small enough
    that estimation with HMC on the tangent plane (red) is sufficient. 
    However, in the region $\mu \in \left[2,3\right]$ an ML HMC (blue) is used 
    to narrow particular values for which the sign problem becomes untraceable.
    The features at $\mu = 0$ and $\mu \to \infty$ are captured as expected.
    Finding the charge plateaus at $\rho(\mu)\sim n$, however, is yet unavailable 
    due to the small $\beta$.
    The dashed line at $\rho(\mu)=\nicefrac{4}{18}$ 
    indicate an expected plateau which may be surmised around $\mu \approx 1$.
}
\label{fig:ChargeDensityScan}
\end{figure}
\begin{figure}
    \centering
\includegraphics[width = 0.9\linewidth,height = 0.3\textheight]{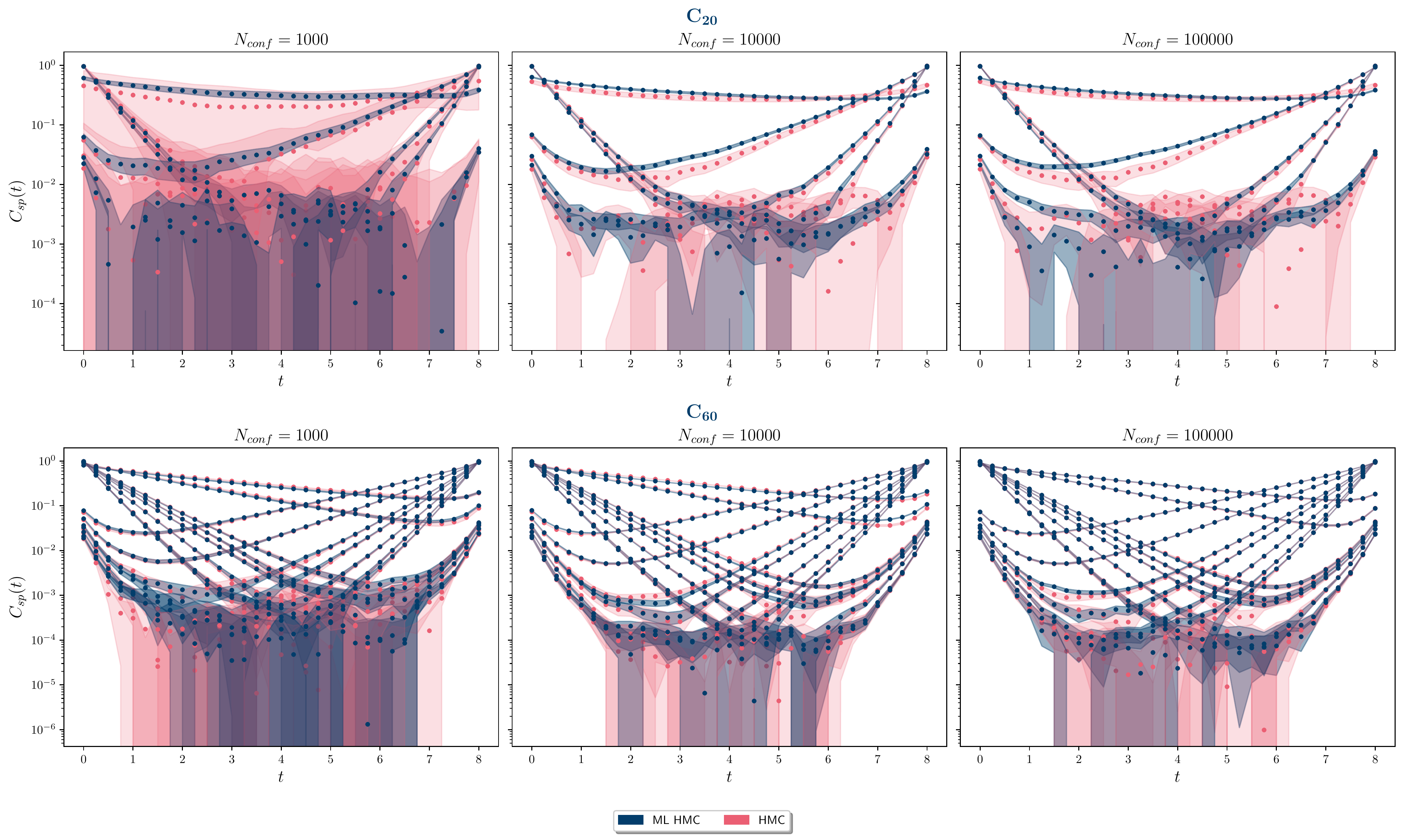}
\caption{
    The correlators of Fullerene $C_{20}$ (upper row) and $C_{60}$ (lower row) are shown.
    A real plane (red) and a tangent plane (blue) HMC at 
    inverse temperature $\beta = 8$ and $\Nt = 32$ time slices are compared.
    We consider an on-site interaction $U = 3$ and zero chemical potential.
    The, here negligible, sign problem solely stems from the non-bipartiteness 
    of the system due to the pentagonal rings.  
    Already, at small number of configurations $\Ncfg \leq \num{10000}$ the 
    signals are very good.
}\label{fig:buckyball-correlators}
\end{figure}
In figure~\ref{fig:SmallSystemCorrelators} correlators for systems with 
2 and 8 sites are compared using the different algorithms HMC (red) --- on the tangent plane --- 
ML HMC (blue), applying the coupling network $\NN$, and exact diagonalization (black)~\cite{rodekampMitigating}.
We use $\Ncfg\in\{\num{1000},\num{100000}\}$ to portray the effect of the 
statistical power on the effective number of configurations. 
Corresponding statistical powers $\abs{\Sigma}$ can be found in~\cite{rodekampMitigating}.
The system parameters --- $\Nt = 32$, $\beta = 4$, $U = 4$ and $\mu = 3$ --- are kept fix.
The ML HMC outperforms the HMC in terms of signal.
The 8 site system has a stronger sign problem to an extend that HMC retrieves no signal.
If a signal is obtained, both algorithms agree with the exact diagonalization.
In figure~\ref{fig:18SiteCorrelator} the correlators for the system with 18 sites are displayed~\cite{rodekampMitigating}.
The system is computed at $\Nt = 32$, $\beta = 4$, $U=3$ and $\mu = 3$.
The sign problem is much stronger than in the previous cases due to the larger volume.
Nevertheless the ML HMC extracts a good signal for the correlators. 
Similar to the 8 site case HMC can't keep up.
Due to the number of sites exact diagonalization is not feasible. 

Continuing the 18 site model --- with $U = 2$, $\beta = 5$, $\Nt=32$ ---
we want to study the charge density~\eqref{eq:charge-density} subjected to the chemical potential. 
This can be seen in figure~\ref{fig:ChargeDensityScan}~\cite{christophThesis}.
We compare tangent plane HMC (red) and ML HMC (blue) using 
the SHIFT network.
Varying the chemical potential has shown that for small and large values 
the sign problem is mild (on the tangent plane).
However, in the intermediate range of $\mu \in [2,3]$ the tangent plane is 
not sufficient for a reasonable estimate, where we apply ML HMC with the SHIFT network.
The point at $\mu = 2.5$ requires more attention and we plan to address it
with the coupling network in the future expecting much better results.
The expected behaviour of the charge density, $\rho(\mu=0) = 0$ and $\rho(\mu\to\infty) \to 1$, 
is found numerically.
The dashed line exemplarily indicates an expected plateau at $\expval{\rho(\mu)} = \nicefrac{4}{18}$.
As it can be seen, this plateau is not fully deducible but may be surmised around $\mu \approx \num{1}$.
Studies of smaller systems, see~\cite{christophThesis}, indicate increasing $\beta$
makes these plateaus more pronounced. %

To probe our method in physically more relevant systems than the 18 Site Graphene sheet, 
we started an investigation of Fullerene $C_{20}$ and $C_{60}$.
The correlators at $\Nt = 32$, $\beta = 8$, $U = 3$ and zero chemical potential 
are displayed in figure~\ref{fig:buckyball-correlators}.
The mild sign problem is solely due to the non-bipartiteness of the lattice structure.
We compare standard HMC (red) with tangent plane HMC (blue) to show 
that the tangent plane obtains a good signal already at small 
number of configurations $\Ncfg = 1000$ in both systems.
For $C_{60}$ the sign problem is negligible and the real plane HMC gives a
good signal too.
For finite chemical potential the sign problem
aggravates as it imposes a second source. 
We are currently working on this particular lattice geometry.

\section{Conclusions}
Simulating systems with strongly correlated electrons is a rather challenging 
task due to the innate sign problem for doped systems.
Current methods, like deformation onto Lefschetz thimbles, suffer from a very difficult 
scaling in computational cost.
We overcome this issue by identifying efficient Neural Network architectures and incorporating them
in a HMC algorithm.
We present first studies of doped Graphene sheets using this enhanced HMC and 
demonstrate a substantial improvement of the signal, effectively mitigating the sign problem.
Considering systems with increasing volume illustrates some stability of this 
method for larger volumes.
Further, preliminary simulation of Fullerene $C_{20}$ and $C_{60}$ at 
vanishing chemical potential are shown.
In the near future we will apply the neural network enhanced HMC also
at finite chemical potential.

\begin{acknowledgments}
We thank the original authors of our recent papers for many helpful discussions and hard work.
This work was funded in part by the NSFC and the Deutsche Forschungsgemeinschaft (DFG, German Research
Foundation) through the funds provided to the Sino-German Collaborative
Research Center ``Symmetries and the Emergence of Structure in QCD''
(NSFC Grant No.~12070131001, DFG Project-ID 196253076 -- TRR110)
as well as the STFC Consolidated Grant ST/T000988/1.  
MR was supported under the RWTH Exploratory Research Space (ERS) grant PF-JARA-SDS005.
We gratefully acknowledge the computing time granted by the JARA Vergabegremium and provided on the JARA Partition part of the supercomputer JURECA at Forschungszentrum Jülich.
\end{acknowledgments}
 
\bibliographystyle{apsrev4-1}

\end{document}